# Phase Transitions in Low-Dimensional Layered Double Perovskites: The Role of the Organic Moieties

*Beatriz Martín-García* [†,ϒ,‡,*], *Davide Spirito* [†,Θ,‡,*], *Giulia Biffi* [†,∥], *Sergey Artyukhin* [†], *Francesco Bonaccorso* [†,#], *Roman Krahne* [†].

[†]Istituto Italiano di Tecnologia, Via Morego 30, 16163 Genova (Italy)

[ϒ]CIC nanoGUNE, Tolosa Hiribidea, 76, 20018 Donostia-San Sebastian, Basque Country (Spain)

[Θ]IHP–Leibniz-Institut für innovative Mikroelektronik, Im Technologiepark 25, 15236 Frankfurt (Oder) (Germany)

[∥]Dipartimento di Chimica e Chimica Industriale, Università degli Studi di Genova, Via Dodecaneso 31, 16146 Genova (Italy)

[#]BeDimensional S.p.A., Via Lungotorrente secca 3d, 16163 Genova (Italy)

AUTHOR INFORMATION

**Corresponding Author**

*Beatriz Martín-García (b.martingarcia@nanogune.eu) and Davide Spirito (spirito@ihp-microelectronics.com)

ABSTRACT. Halide double perovskites are an interesting alternative to Pb-containing counterparts as active materials in optoelectronic devices. Low-dimensional double perovskites are fabricated by introducing large organic cations, resulting in organic/inorganic architectures with one or more inorganic octahedral layers separated by organic cations. Here, we synthesize layered double perovskites based on 3D $Cs_2AgBiBr_6$ that consist of double (2L) or single (1L) inorganic octahedral layers, using ammonium cations of different size and chemical structure. Temperature-dependent Raman spectroscopy reveals phase transition signatures in both inorganic lattice and organic moieties by detecting variations in their vibrational modes. Changes in the conformational arrangement of the organic cations to an ordered state coincide with a phase transition in the 1L systems with the shortest ammonium moieties. Significant changes of photoluminescence intensity observed around the transition temperature suggest that optical properties may be deeply affected by the octahedral tilts emerging at the phase transition.

**TOC GRAPHICS**

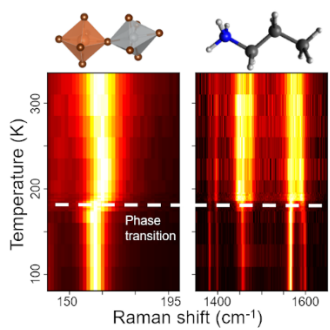

In the last years, metal-halide perovskites have attracted much interest for the development of next-generation optoelectronic devices due to their outstanding figures of merit in photovoltaics, exceeding 23% power conversion efficiency (PCE),[1] and in light emitting devices (LEDs), with nearly 100% internal quantum efficiency,[2] together with defect tolerance, low-cost solution processing and tunable emission across the visible spectrum.[3–5] However, these devices suffer from severe limitations, such as low ambient stability,[4] ion migration under operation,[4,6] and the presence of toxic Pb.[4] In this respect, the substitution of $Pb^{2+}$ by the combination of trivalent ($Bi^{3+}$, $Sb^{3+}$) and monovalent ($Ag^+$, $Cu^+$, $Au^+$, $K^+$) cations leads to the formation of double perovskites, which demonstrated ambient stability, but inferior device performance compared to Pb-halide perovskites.[4,7,8] The $Cs_2AgBiBr_6$ double perovskite has demonstrated good performance in solar cells reaching a PCE 2.84%,[9–15] and in photodetectors with a high detectivity of $3.29 \times 10^{12}$ Jones and fast response of 17 ns,[16] as well as application in X-Ray detectors[17] and memristors.[18] The optoelectronic properties of $Cs_2AgBiBr_6$ stimulated the development of low-dimensional (2D) perovskites from their 3D counterparts by the introduction of large A-site cations. These layered double perovskites feature many interesting properties, for example the transition from an indirect to a direct band gap material by tuning the number of adjacent octahedra layers.[19–22] However, for their further development, it is essential to understand how the crystal structure and lattice dynamics affect the optical and electronic properties, and their relation to electron-phonon coupling and phase transitions.[23–25]

In this work, we study layered double perovskites derived from $Cs_2AgBiBr_6$ focusing on the influence of the organic cation layer on the optical and vibrational properties. We use a series of alkyl and phenyl ammonium cations with varying size and structure, and grow crystals with single (1L) and double (2L) octahedra layers. Temperature-dependent micro-Raman spectroscopy allows to determine the presence or absence of phase transitions. We show that the intermolecular interactions between the ammonium alkyl-chain cations and the inorganic octahedral layer are closely related to phase transitions in the 1L layered crystal structures. Concerning the emission properties, the photoluminescence (PL) spectra strongly depend on the number of adjacent octahedral layers in the crystal structure as well as on a structural phase transition. This is in agreement with the transition from an indirect band gap for the 3D crystal[21,23,25–28] to a direct band gap for the 1L system[19]. The influence of the octahedral distortions on the optical properties may be enhanced by an extremely flat conduction band (CB) in 1L compounds, such that minor structural changes are sufficient to affect them. Our work provides novel insights into the correlation of structural and optical properties in layered double perovskites, and demonstrates how the proper selection of the organic cations can be used to avoid or deliberately trigger phase transitions, with application in devices for improved thermal stability,[14] or exploiting switching.[29]

We synthesized layered double perovskite crystals derived from the $Cs_2AgBiBr_6$ (3D) system using revised recipes[19,30]. To evaluate the effect of length and stereochemistry of the organic moieties on the interaction between octahedra layers, we selected a series of alkyl cations and a phenyl cation, concretely: propylammonium (PA), butylammonium (BA), phenylethylammonium (PEA), decylammonium (DA) and dodecylammonium (DoA) (see Figure 1a-g and Supporting Information (SI), Experimental methods section and Figure S1). By selecting the concentrations of the organic moieties and inorganic precursors ($BiBr_3$, $AgBr$ and $CsBr$), we obtained (R-$NH_3^+$)$_4AgBiBr_8$ crystals that consists of single octahedra layers separated by organic cations (hereafter referred to as 1L), and (R-$NH_3^+$)$_2CsAgBiBr_7$ crystals with two adjacent octahedra layers enabled by the presence of $Cs^+$ (noted as 2L) (see Figure 1d-f).[19] X-ray diffraction (XRD) spectra of the different samples are shown in Figure S2 and allow for the 1L systems to directly evaluate the inter-distance between the octahedra layers (*d*) (Figure 1h). Here, the *d* value increases with the length of the molecules up to 20-23 Å, after which the alkyl chains start to cross or entangle as depicted in Figure S3, in agreement with literature[19,20].

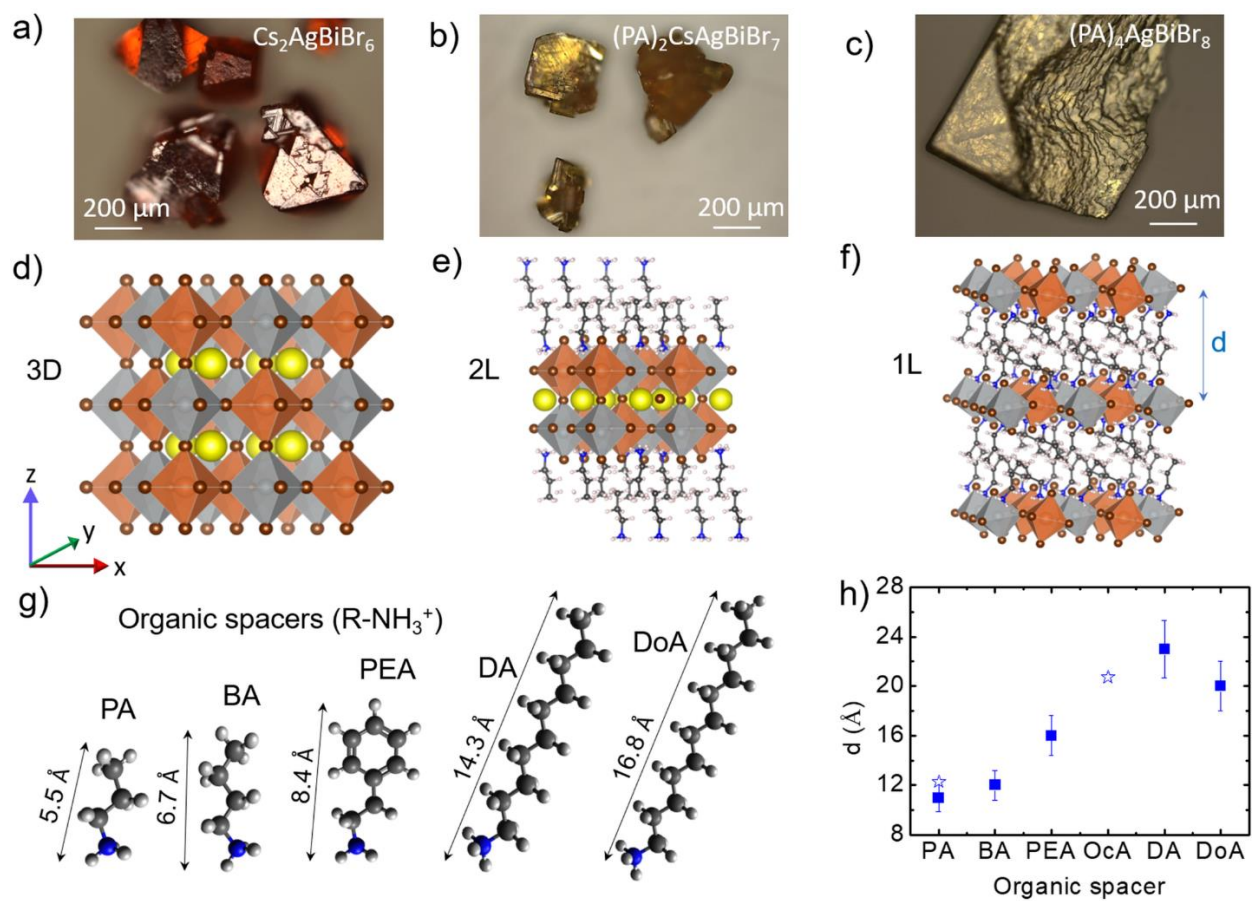

**Figure 1.** (a-c) Representative optical microscope images of the 3D $Cs_2AgBiBr_6$ crystals, and $(PA)_2CsAgBiBr_7$ (2L) and $(PA)_4AgBiBr_8$ (1L) layered perovskites. (d-f) Crystal structures of the 3D, 2L-PA and 1L-PA samples, drawn using VESTA 3 software[31] based on the crystallographic data from refs[19,20]. In (c) the octahedra layers inter-distance is indicated as *d*. (g) Schemes of the ammonium cations used in this work and their length estimated using Avogadro software.[32] (h) Inter-distance between the octahedra layers (*d*) for the 1L samples with the different organic

moieties obtained from XRD analysis. Additionally, values from literature[20] are displayed with stars for PA and octylammonium (OcA) perovskites.

The Raman spectra of the 3D, PA, BA, PEA and DA systems recorded at 80K are shown in Figure 2a in the 100-300 cm$^{-1}$ spectral range. All spectra show a dominant peak in the range around 170 cm$^{-1}$ that is assigned to the $A_{1g}$ mode[33] related to stretching of the $(AgBr_6)^{5-}$ and $(BiBr_6)^{3-}$ octahedra.[23,34] The $A_{1g}$ is also associated to the longitudinal-optical (LO) phonon of the inorganic lattice,[34–36] and occurs at the highest frequency (178 cm$^{-1}$) for the 3D crystal, due to its higher stiffness compared to the 2L and 1L crystals (in this order), which is reasonable due to the fully inorganic structure in all three spatial dimensions. Indeed, compared to the 3D crystal, the introduction of the organic layers leads to a red shift of the $A_{1g}$ mode (Figure 2b), which is clearly more pronounced for the 1L samples, in which the organic layers are in contact with all octahedra (such behavior is also observed at room temperature, see Figure S5a-b). Interestingly, the $A_{1g}$ mode for the PEA system does not follow this trend, since here the red shift is significant already for the 2L structure. This difference could be attributed to the different stereochemistry and arrangement of the phenyl rings, compared to the alkyl ammonium moieties, that result in a $\pi - \pi$ stacking in the organic layer.[37,38] The deviation of $A_{1g}$ mode in 1L-DA sample is possibly due to alkyl chains entanglement causing higher stiffness than in the other 1L-crystals. Furthermore, distortions and octahedral tilts in the inorganic layer can affect the phonon modes, as will be discussed later on. For the layered systems a second, much weaker, peak can be identified around 140 cm$^{-1}$ that we assign to a vibrational mode with $E_g$ symmetry.[23,34] Additionally we do not observe differences in the Raman spectrum when the material is mechanically exfoliated in form of flakes compared to the bulk crystal (Figure S5c).

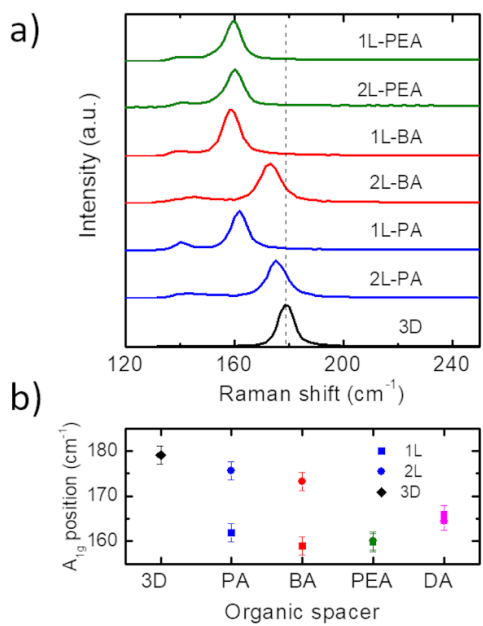

**Figure 2.** (a) Raman spectra recorded with a micro-setup in the spectral range of the $A_{1g}$ mode from the 3D crystal and layered systems with PA, BA and PEA organic moieties. Temperature was 80K, and the laser excitation wavelength was 532 nm. (b) Spectral position of the $A_{1g}$ mode for the different samples.

Temperature-dependent Raman data for the 3D, BA, PA, and PEA systems are shown in Figure 3 (see Figure S6-S9 for additional data). For the 3D crystal, a careful inspection of the $A_{1g}$ Raman band reveals a slight change of the mode frequency (~2 cm$^{-1}$ at 125-150K, accompanied by a peak broadening, Figure S6). This behavior can be ascribed to a phase transition from tetragonal to cubic structure,[25] also observed for other double perovskites.[39] Noteworthy, for the 1L systems the temperature-dependent behavior is strongly influenced by the type of the organic moiety, and shows abrupt shifts at well-defined temperatures (indicated by the white dashed lines) for the 1L-PA (~172K) and 1L-BA (~282K) samples. For both samples we observe a blue shift of the $A_{1g}$ mode, accompanied by a sudden broadening, with increasing temperature in this region. Comparison of the $A_{1g}$ Raman band with the vibrational bands of the molecules in the 1100-1700

cm$^{-1}$ range around the transition temperature reveals a concomitant abrupt broadening of the CH$_2$ and NH$_3$ vibrational bands[40,41] (Figure S10-S11). This correlation demonstrates a stringent relation of the conformational arrangement of the molecules in the organic layer to the vibrational modes of the inorganic octahedra. These primary amines are known to undergo a transition from the solid to the liquid phase at 200K for PA and at 224K for BA.[42] This is in good agreement with our observation that the phase-transition occurs at higher temperatures in 1L-BA than in 1L-PA. We can therefore conclude that the transition in the organic layer triggers a structural change in the octahedra layer that we assign to modifications in the octahedral tilt pattern, as reported in ref.[19]. Illustrations of such different tilt patterns are depicted in Figure S4. Furthermore, we note that the arrangement of the disordered phase in the organic layer, which correlates with a marked tilting of the octahedra, results in a blue shift of the A$_{1g}$ mode compared to the ordered phase, where there is a less distorted octahedra lattice. Similarly, the red shift of the A$_{1g}$ mode in the low-dimensional systems (with respect to the frequency of the 3D system – see Figure 2) reflects how the organic layer acts on the inorganic octahedra lattice.

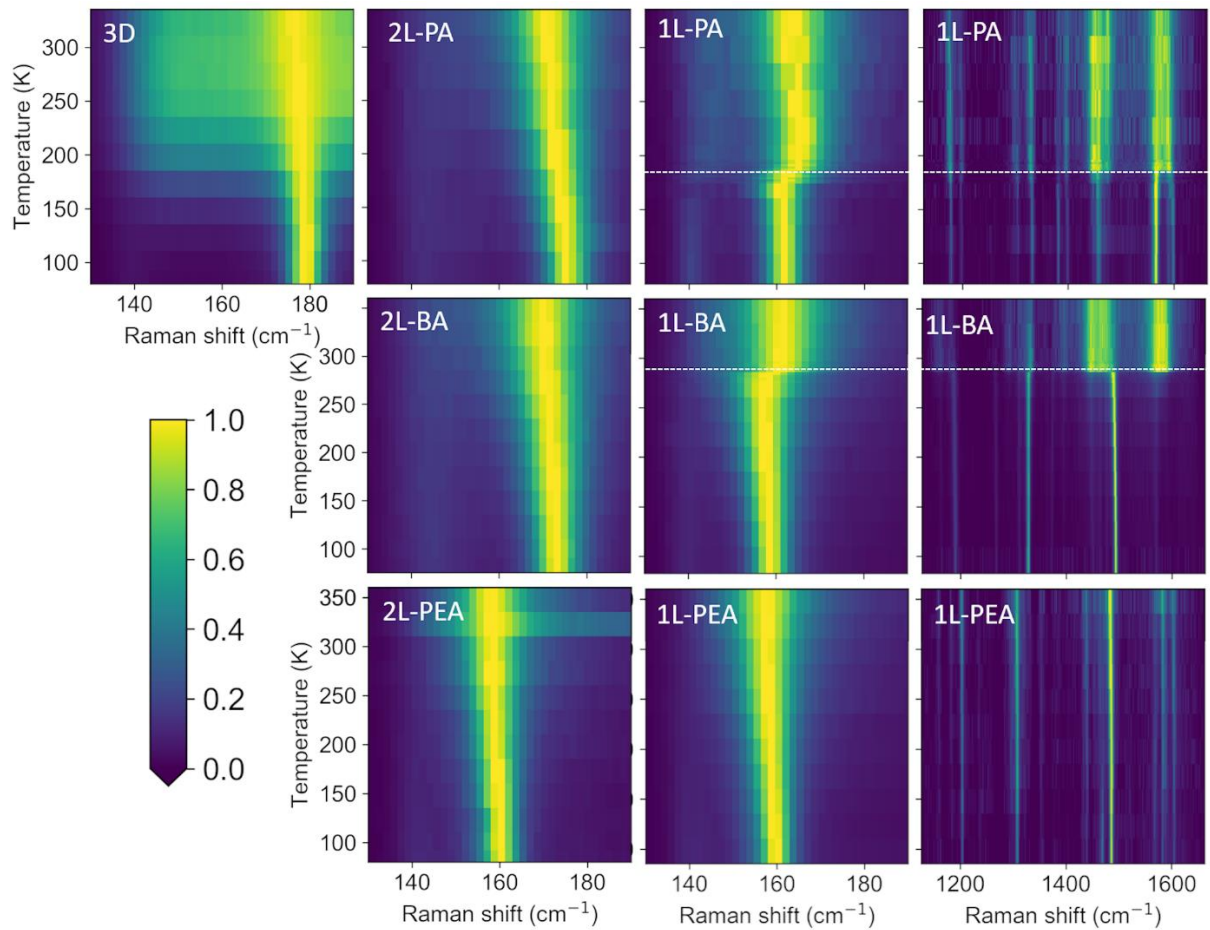

**Figure 3.** Temperature-dependent Raman spectra in the range of the $A_{1g}$ band for the PA, BA and PEA systems and the 3D crystal. For the 1L systems also spectra in the range from 1100-1700 cm$^{-1}$ are shown, in which the vibrational resonances of the organic cations occur. The horizontal dashed lines indicate the temperature in which the $A_{1g}$ mode in the 1L-PA and 1L-BA samples undergoes a sudden shift, coinciding with a drastic change in the width of the molecular vibrational modes around 1200 cm$^{-1}$ (C-N stretching), 1340 cm$^{-1}$ (CH$_2$ twisting/wagging type), 1500 cm$^{-1}$ (CH$_2$ bending) and 1600 cm$^{-1}$ (NH$_3^+$ bending).[40,41] For the PEA system, no shift of the $A_{1g}$ mode or linewidth broadening of the Raman modes related to molecular vibrations are observed.

The distortion of the octahedra lattice significantly affects the electronic structure and the transition from an indirect to a direct band gap, and therefore is of paramount importance for the optical properties. We quantify the lattice distortion from the corresponding crystallographic data by the parameter $\lambda_{Ag\text{-}Br}$ that expresses the deviation of the Ag-Br bond length compared to the undistorted Ag-Br octahedra lattice (see details in the SI and Table S3). This parameter $\lambda_{Ag\text{-}Br}$ increases from 3D to 2L and from 2L to 1L, and depends on the type of the organic moiety.

To relate the distortions and the vibrational modes to the optical properties and the electronic band structure, we performed temperature-dependent absorbance and PL measurements, and band structure calculations using density-functional theory (DFT) for different systems. In the absorbance spectra recorded from 3D, 2L-PA, and 1L-PA samples at different temperatures (Figure 4a, and Figure S12-S14 and Table S4 for additional data), we do not find any significant shift of the absorption peak related to the direct bandgap,[27] but only a slight broadening with increasing temperature, in agreement with literature.[19] However, the crystal structure of the samples has a strong impact on the position of the absorption peak that blue-shifts when passing from 3D to 2L to 1L (Figure S13).

Concerning the PL properties shown in Figure 4b (see Figures S15-16 for additional data), for the 3D sample at 80K, we observe a broad emission peak centered ~2 eV, and similarly for the 2L samples, as reported in literature[19,27,28]. Although this emission was firstly related to color centers,[21,23,25] recent reports support an indirect exciton recombination, which could be accompanied by a weak blue PL emission close to the absorption peak position,[27,28,43] and in agreement with the strong electron-phonon coupling found in these materials[23,34] (see Table S5). For the 1L samples, a broad and asymmetric peak with a maximum at ~3 eV is observed, similar to what has been reported for 1L-BA at 80K,[19] and ascribed to fast recombination at the direct band edge.[19,27,28]

Regarding the temperature dependence (Figure 4c), the PL intensity of the 3D, 2L-PA and 1L-PA samples strongly increases below the corresponding phase transition temperatures (122K,[25] 222K[44] and 172K, respectively). However, the change in the PL intensity around the phase transition is more marked in the case of the 1L-PA, with an increase of ~95% in 15K, while below the transition temperature is relatively stable with only a minor decrease (~12% in 42K). This could be attributed to the conformational changes of the organic moieties detected by Raman spectroscopy (Figure 3), which can affect the band structure as extracted from our DFT calculations discussed below.

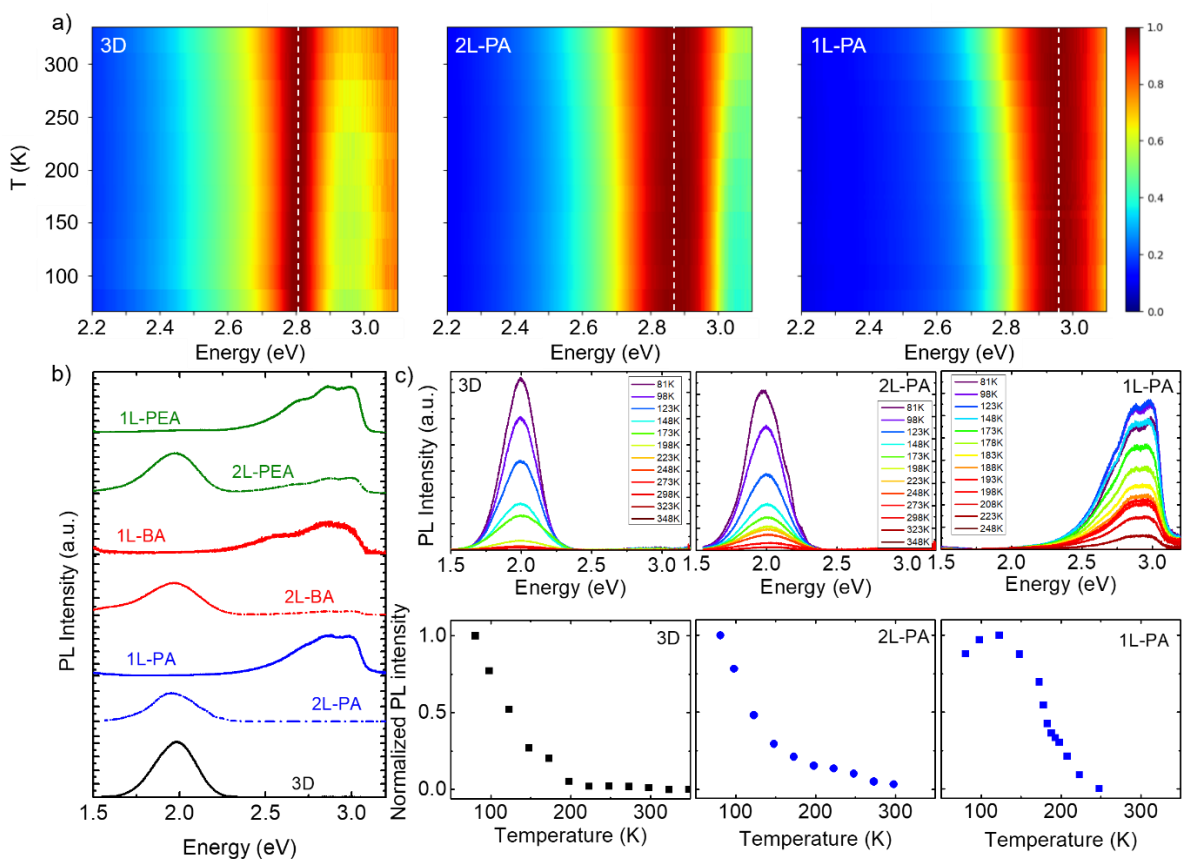

**Figure 4.** (a) Normalized spectral intensity maps of the temperature-dependent absorbance for 3D, 2L and 1L PA-films. The dashed line indicates the position of the absorption peak. (b) Photoluminescence spectra of the 3D, and the PA, BA, and PEA systems recorded from the corresponding crystals at 80K and laser excitation wavelength at 325 nm. (c) Temperature-dependent PL spectra of the 3D, 2L-PA and 1L-crystals (top), and the evolution of the PL intensity (bottom).

To gain deeper understanding of the influence of the phase transition on the optical properties, we calculate the band structure[45,46] for the 1L-BA and 2L-BA crystals above and below the transition temperature, including 1L-PA and 2L-PA structures at room temperature to evaluate the effect of the organic cation. Our calculations revealed a strong influence of spin-orbit coupling (Figure S17), inducing a splitting of 1 eV of the lowest CB. The analysis of low- and room-temperature band structures was performed for 1L-BA and 2L-BA systems (Figure 5a-c). Similar behaviors can be expected at low-temperatures for the 1L-PA and 1L-BA systems, since the band structures of 1L-PA and 1L-BA at room-temperature (Figure 5c,d) are almost identical within a 5 eV energy window around the Fermi level. This is in line with the absence of organic states close to the Fermi level, the only difference in the two compounds being an extra $CH_2$ group on the alkyl chain. Both 1L-PA and 1L-BA are likely to have similar distortions of the inorganic framework, which is corroborated by additional calculations on the 2L-BA and 2L-PA structures (Figure S18).

The calculated bands and density of states of 1L structures are in agreement with those reported in literature[19], showing a direct band gap at the Γ point with Br-*4p* and Ag-*4d* orbitals forming the top valence band and Bi-*6p* and Br-*4p* orbitals forming the CBs. Consistently with Ref.[19], an indirect band gap is obtained for the 2L-system (Figure 5b and Figure S18).

The 1L-BA structure features different octahedral tilt patterns at low- and room-temperatures. At 298K the octahedral tilts are within *xz* plane ($a^0b^+c^0$ in Glazer's notation[47]), while below the transition temperature a tilt in the *xy* plane appears ($a^0a^0c^-$). A striking feature of 1L structures is a rather flat bottom CB, which is related to the strong decrease of Bi-Bi hopping integral due to energetically well-separated Ag orbitals within the checkerboard ordering of the metal ions (Figure S19a). Indeed, in 2L structures with shorter Ag-Br bonds along the *c* axis and higher number of connected octahedra than the 1L systems, the dispersion of the lowest CB is much larger (~0.5 eV). The increased amplitude of the $a^0b^+c^0$ octahedral tilt at room temperature causes the splitting of 0.1 eV of the lowest conduction and the highest valence bands. Combined with the flatness of the bands, this makes possible for small distortions to produce significant changes in the band structure across the bandgap, thus affecting the optical properties. This could be at the origin of the change in PL intensity observed around the phase transition temperature for the 1L-PA sample. Since the CB is essentially flat, the transition energy remains almost unchanged, in good agreement with our absorption and PL results. Additionally, thermally excited polar modes could cause a shift in the CB minimum *via* the Rashba effect, leading to indirect bandgap (discussion in Figure S19b). However, these modes are not visible in the centrosymmetric 1L crystal structures experimentally determined[19,21].

Overall, the analysis of the band structures of 1L-BA system above and below the phase transition temperature shows how variations in octahedral tilt patterns mostly affect the flat CB, opening interesting perspectives on the effects of electron-phonon coupling on these materials.

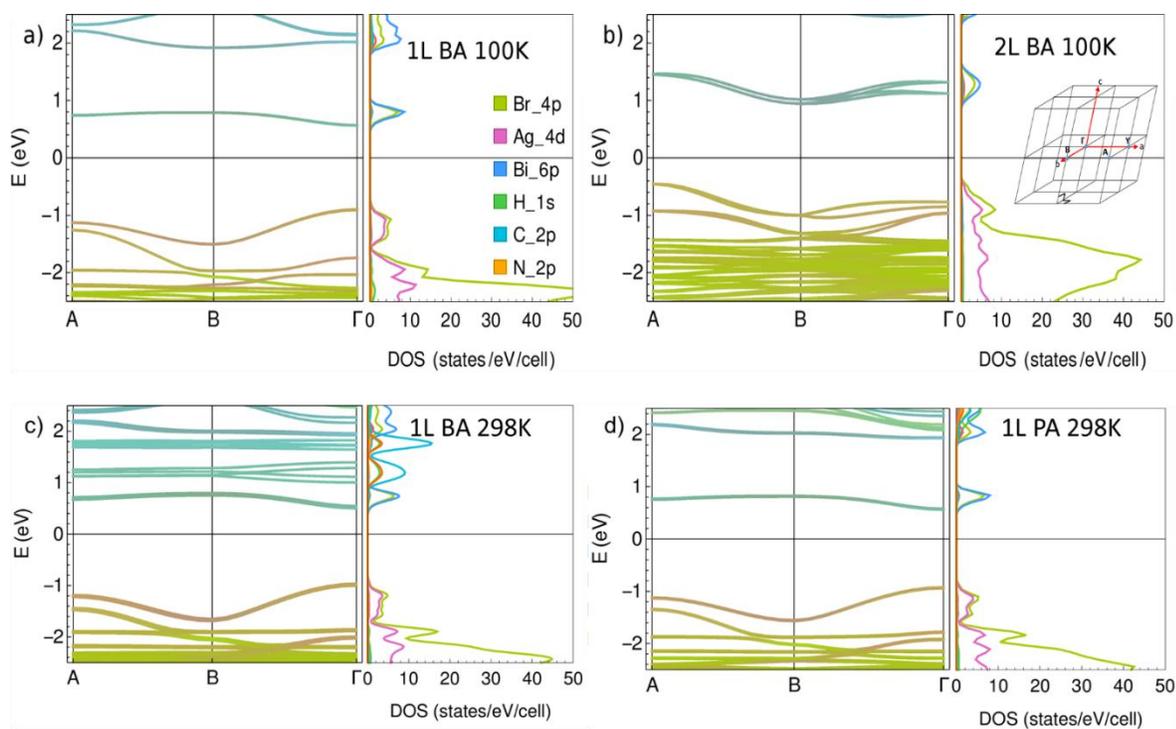

**Figure 5.** Band structure and projected density of states of (a) 1L-BA at 100K, (b) 2L-BA at 100K, (c) 1L-BA at 298K and (d) 1L-PA at 298K. The colors encode the contribution of corresponding atomic orbitals to the states at each *k*-point. Inset: (b) the Brillouin zone and high-symmetry points used in the band structure plots.

In summary, we performed a systematic study of layered crystals derived from the $Cs_2AgBiBr_6$ double perovskite, consisting of single layers (1L) or bilayers (2L) of inorganic octahedra with organic moieties of different length and structure. Our work elucidates the interplay of lattice vibrations, crystal phases and optical properties combining temperature-dependent Raman, absorbance, and photoluminescence spectroscopy with DFT modeling. Here we show that changes in the conformational rearrangement, from a *"solid-like"* phase (ordered) to *"liquid-like"* (disordered) one, of the organic cations can translate to phase transitions in the layered crystals,

  

and that such transitions influence the PL intensity, but not the emission wavelength. Our findings open pathways to manipulate this organic-mediated phase transition by properly selecting the ammonium moieties. The possibility to independently tune the band gap (*via* the composition of the octahedral layers) and structural phase transitions (*via* the choice of organic moieties) in these materials is appealing towards applications in optoelectronic devices such as solar cells or LEDs in terms of thermal stability. In addition, these properties can be exploited in devices using the phase transition for active switching.

ASSOCIATED CONTENT

**Supporting Information**. Experimental methods; Optical microscope images; XRD patterns and analysis; Raman spectra at room temperature; detailed Raman spectroscopy characterization; additional absorbance and PL spectroscopy characterization; and Computational details.

AUTHOR INFORMATION

**Corresponding Author**

*Beatriz Martín-García (b.martingarcia@nanogune.eu) and Davide Spirito (spirito@ihp-microelectronics.com)

**Notes**

The manuscript was written through contributions of all authors. All authors have given approval to the final version of the manuscript. [‡]The authors B.M-G. and D.S. contributed equally. G.B. and S.A. performed the DFT modelling. The authors declare no competing financial interest.

ACKNOWLEDGMENT

B.M-G. and F.B. acknowledge the funding from the European Union's Horizon 2020 research and innovation programme under grant agreement no. 785219 (GrapheneCore2). B.M-G. thanks also to Gipuzkoa Council (Spain) in the frame of Gipuzkoa Fellows Program. D.S. acknowledges the funding from the German Research Foundation (DFG) within the project ESSENCE. This work is supported by the Spanish MINECO under the María de Maeztu Units of Excellence Programme (MDM-2016-0618). We thank Prof. G. Capellini (IHP) for fruitful discussions regarding the experimental spectroscopic set-up and Dr. D. Kot (IHP) for the assistance with the temperature-dependent absorbance set-up. We acknowledge Dr. M. Prato for the access to the Materials Characterization Facility – IIT for the XRD characterization, and Dr. A. Toma for the access and E. Rondanina for her assistance in the laser-writer patterning of markers on Si/SiO$_2$ substrates at the Clean Room Facility – IIT.

REFERENCES

(1) NREL. Best Research-Cell Efficiency Chart https://www.nrel.gov/pv/cell-efficiency.html (accessed Mar 25, 2020).
(2) Zhao, B.; Bai, S.; Kim, V.; Lamboll, R.; Shivanna, R.; Auras, F.; Richter, J. M.; Yang, L.; Dai, L.; Alsari, M.; et al. High-Efficiency Perovskite–Polymer Bulk Heterostructure Light-Emitting Diodes. *Nat. Photonics* **2018**, *12* (12), 783–789. https://doi.org/10.1038/s41566-018-0283-4.
(3) Grancini, G.; Nazeeruddin, M. K. Dimensional Tailoring of Hybrid Perovskites for Photovoltaics. *Nat. Rev. Mater.* **2019**, *4* (1), 4–22. https://doi.org/10.1038/s41578-018-0065-0.
(4) Jena, A. K.; Kulkarni, A.; Miyasaka, T. Halide Perovskite Photovoltaics: Background, Status, and Future Prospects. *Chem. Rev.* **2019**, *119* (5), 3036–3103. https://doi.org/10.1021/acs.chemrev.8b00539.
(5) Smith, M. D.; Connor, B. A.; Karunadasa, H. I. Tuning the Luminescence of Layered Halide Perovskites. *Chem. Rev.* **2019**, *119* (5), 3104–3139. https://doi.org/10.1021/acs.chemrev.8b00477.
(6) Capasso, A.; Matteocci, F.; Najafi, L.; Prato, M.; Buha, J.; Cinà, L.; Pellegrini, V.; Carlo, A. D.; Bonaccorso, F. Few-Layer MoS$_2$ Flakes as Active Buffer Layer for Stable Perovskite Solar Cells. *Adv Energy Mater* **2016**, *6* (16), 1600920. https://doi.org/10.1002/aenm.201600920.
(7) Xiao, Z.; Song, Z.; Yan, Y. From Lead Halide Perovskites to Lead-Free Metal Halide Perovskites and Perovskite Derivatives. *Adv. Mater.* **2019**, 1803792. https://doi.org/10.1002/adma.201803792.

(8) Vargas, B.; Rodríguez-López, G.; Solis-Ibarra, D. The Emergence of Halide Layered Double Perovskites. *ACS Energy Lett.* **2020**, *5* (11), 3591–3608. https://doi.org/10.1021/acsenergylett.0c01867.

(9) Greul, E.; Petrus, M. L.; Binek, A.; Docampo, P.; Bein, T. Highly Stable, Phase Pure Cs$_2$AgBiBr$_6$ Double Perovskite Thin Films for Optoelectronic Applications. *J. Mater. Chem. A* **2017**, *5* (37), 19972–19981. https://doi.org/10.1039/C7TA06816F.

(10) Wang, M.; Zeng, P.; Bai, S.; Gu, J.; Li, F.; Yang, Z.; Liu, M. High-Quality Sequential-Vapor-Deposited Cs$_2$AgBiBr$_6$ Thin Films for Lead-Free Perovskite Solar Cells. *Sol. RRL* **2018**, *2* (12), 1800217. https://doi.org/10.1002/solr.201800217.

(11) Pantaler, M.; Cho, K. T.; Queloz, V. I. E.; García Benito, I.; Fettkenhauer, C.; Anusca, I.; Nazeeruddin, M. K.; Lupascu, D. C.; Grancini, G. Hysteresis-Free Lead-Free Double-Perovskite Solar Cells by Interface Engineering. *ACS Energy Lett.* **2018**, *3* (8), 1781–1786. https://doi.org/10.1021/acsenergylett.8b00871.

(12) Igbari, F.; Wang, R.; Wang, Z.-K.; Ma, X.-J.; Wang, Q.; Wang, K.-L.; Zhang, Y.; Liao, L.-S.; Yang, Y. Composition Stoichiometry of Cs$_2$AgBiBr$_6$ Films for Highly Efficient Lead-Free Perovskite Solar Cells. *Nano Lett.* **2019**, *19* (3), 2066–2073. https://doi.org/10.1021/acs.nanolett.9b00238.

(13) Zhang, Z.; Wu, C.; Wang, D.; Liu, G.; Zhang, Q.; Luo, W.; Qi, X.; Guo, X.; Zhang, Y.; Lao, Y.; et al. Improvement of Cs$_2$AgBiBr$_6$ Double Perovskite Solar Cell by Rubidium Doping. *Org. Electron.* **2019**, *74*, 204–210. https://doi.org/10.1016/j.orgel.2019.06.037.

(14) Wu, C.; Zhang, Q.; Liu, Y.; Luo, W.; Guo, X.; Huang, Z.; Ting, H.; Sun, W.; Zhong, X.; Wei, S.; et al. The Dawn of Lead-Free Perovskite Solar Cell: Highly Stable Double Perovskite Cs$_2$AgBiBr$_6$ Film. *Adv. Sci.* **2018**, *5* (3), 1700759. https://doi.org/10.1002/advs.201700759.

(15) Yang, X.; Wang, W.; Ran, R.; Zhou, W.; Shao, Z. Recent Advances in Cs$_2$AgBiBr$_6$ -Based Halide Double Perovskites as Lead-Free and Inorganic Light Absorbers for Perovskite Solar Cells. *Energy Fuels* **2020**, *34* (9), 10513–10528. https://doi.org/10.1021/acs.energyfuels.0c02236.

(16) Yang, J.; Bao, C.; Ning, W.; Wu, B.; Ji, F.; Yan, Z.; Tao, Y.; Liu, J.; Sum, T. C.; Bai, S.; et al. Stable, High-Sensitivity and Fast-Response Photodetectors Based on Lead-Free Cs$_2$AgBiBr$_6$ Double Perovskite Films. *Adv. Opt. Mater.* **2019**, 1801732. https://doi.org/10.1002/adom.201801732.

(17) Pan, W.; Wu, H.; Luo, J.; Deng, Z.; Ge, C.; Chen, C.; Jiang, X.; Yin, W.-J.; Niu, G.; Zhu, L.; et al. Cs$_2$AgBiBr$_6$ Single-Crystal X-Ray Detectors with a Low Detection Limit. *Nat. Photonics* **2017**, *11* (11), 726–732. https://doi.org/10.1038/s41566-017-0012-4.

(18) Cheng, X.; Qian, W.; Wang, J.; Yu, C.; He, J.; Li, H.; Xu, Q.; Chen, D.; Li, N.; Lu, J. Environmentally Robust Memristor Enabled by Lead-Free Double Perovskite for High-Performance Information Storage. *Small* **2019**, *15* (49), 1905731. https://doi.org/10.1002/smll.201905731.

(19) Connor, B. A.; Leppert, L.; Smith, M. D.; Neaton, J. B.; Karunadasa, H. I. Layered Halide Double Perovskites: Dimensional Reduction of Cs$_2$AgBiBr$_6$. *J. Am. Chem. Soc.* **2018**, *140* (15), 5235–5240. https://doi.org/10.1021/jacs.8b01543.

(20) Mao, L.; Teicher, S. M. L.; Stoumpos, C. C.; Kennard, R. M.; DeCrescent, R. A.; Wu, G.; Schuller, J. A.; Chabinyc, M. L.; Cheetham, A. K.; Seshadri, R. Chemical and Structural Diversity of Hybrid Layered Double Perovskite Halides. *J. Am. Chem. Soc.* **2019**, *141* (48), 19099–19109. https://doi.org/10.1021/jacs.9b09945.

(21) Fang, Y.; Zhang, L.; Wu, L.; Yan, J.; Lin, Y.; Wang, K.; Mao, W. L.; Zou, B. Pressure-Induced Emission (PIE) and Phase Transition of a Two-dimensional Halide Double Perovskite (BA)$_4$AgBiBr$_8$ (BA=CH$_3$(CH$_2$)$_3$NH$_3^+$). *Angew. Chem. Int. Ed.* **2019**, *58* (43), 15249–15253. https://doi.org/10.1002/anie.201906311.

(22) McClure, E. T.; McCormick, A. P.; Woodward, P. M. Four Lead-Free Layered Double Perovskites with the *n* = 1 Ruddlesden–Popper Structure. *Inorg. Chem.* **2020**, *59* (9), 6010–6017. https://doi.org/10.1021/acs.inorgchem.0c00009.

(23) Zelewski, S. J.; Urban, J. M.; Surrente, A.; Maude, D. K.; Kuc, A.; Schade, L.; Johnson, R. D.; Dollmann, M.; Nayak, P. K.; Snaith, H. J.; et al. Revealing the Nature of Photoluminescence Emission in the Metal-Halide Double Perovskite Cs$_2$AgBiBr$_6$. *J. Mater. Chem. C* **2019**, *7* (27), 8350–8356. https://doi.org/10.1039/C9TC02402F.

(24) Kentsch, R.; Scholz, M.; Horn, J.; Schlettwein, D.; Oum, K.; Lenzer, T. Exciton Dynamics and Electron–Phonon Coupling Affect the Photovoltaic Performance of the Cs$_2$AgBiBr$_6$ Double Perovskite. *J. Phys. Chem. C* **2018**, *122* (45), 25940–25947. https://doi.org/10.1021/acs.jpcc.8b09911.

(25) Schade, L.; Wright, A. D.; Johnson, R. D.; Dollmann, M.; Wenger, B.; Nayak, P. K.; Prabhakaran, D.; Herz, L. M.; Nicholas, R.; Snaith, H. J.; et al. Structural and Optical Properties of Cs$_2$AgBiBr$_6$ Double Perovskite. *ACS Energy Lett.* **2019**, *4* (1), 299–305. https://doi.org/10.1021/acsenergylett.8b02090.

(26) Su, J.; Mou, T.; Wen, J.; Wang, B. First-Principles Study on the Structure, Electronic, and Optical Properties of Cs$_2$AgBiBr$_{6-x}$Cl$_x$ Mixed-Halide Double Perovskites. *J. Phys. Chem. C* **2020**, *124* (9), 5371–5377. https://doi.org/10.1021/acs.jpcc.9b11827.

(27) Schmitz, A.; Schaberg, L. L.; Sirotinskaya, S.; Pantaler, M.; Lupascu, D. C.; Benson, N.; Bacher, G. Fine Structure of the Optical Absorption Resonance in Cs$_2$AgBiBr$_6$ Double Perovskite Thin Films. *ACS Energy Lett.* **2020**, *5* (2), 559–565. https://doi.org/10.1021/acsenergylett.9b02781.

(28) Dey, A.; Richter, A. F.; Debnath, T.; Huang, H.; Polavarapu, L.; Feldmann, J. Transfer of Direct to Indirect Bound Excitons by Electron Intervalley Scattering in Cs$_2$AgBiBr$_6$ Double Perovskite Nanocrystals. *ACS Nano* **2020**, *14* (5), 5855–5861. https://doi.org/10.1021/acsnano.0c00997.

(29) Zeb, A.; Sun, Z.; Khan, A.; Zhang, S.; Khan, T.; Asghar, M. A.; Luo, J. [C$_6$H$_{14}$N]PbI$_3$: A One-Dimensional Perovskite-like Order–Disorder Phase Transition Material with Semiconducting and Switchable Dielectric Attributes. *Inorg. Chem. Front.* **2018**, *5* (4), 897–902. https://doi.org/10.1039/C7QI00722A.

(30) Slavney, A. H.; Hu, T.; Lindenberg, A. M.; Karunadasa, H. I. A Bismuth-Halide Double Perovskite with Long Carrier Recombination Lifetime for Photovoltaic Applications. *J. Am. Chem. Soc.* **2016**, *138* (7), 2138–2141. https://doi.org/10.1021/jacs.5b13294.

(31) Momma, K.; Izumi, F. *VESTA* 3 for Three-Dimensional Visualization of Crystal, Volumetric and Morphology Data. *J. Appl. Crystallogr.* **2011**, *44* (6), 1272–1276. https://doi.org/10.1107/S0021889811038970.

(32) Hanwell, M. D.; Curtis, D. E.; Lonie, D. C.; Vandermeersch, T.; Zurek, E.; Hutchison, G. R. Avogadro: An Advanced Semantic Chemical Editor, Visualization, and Analysis Platform. *J. Cheminformatics* **2012**, *4* (1), 17. https://doi.org/10.1186/1758-2946-4-17.

(33) To make the assignment easier to follow, we kept the notation A$_{1g}$ throughout the text. However, the A$_{1g}$ mode is from the representation of O$_h$, that refers to the point group of 3D crystal. The derived layered crystals from 3D have a point group C$_{2h}$, and therefore, the A$_{1g}$ mode in $O_h$ corresponds to the $A_g$ mode in $C_{2h}$. In the same way, the $E_g$ mode for $O_h$ in 3D splits in two $A_g$ modes in $C_{2h}$.

(34) Steele, J. A.; Puech, P.; Keshavarz, M.; Yang, R.; Banerjee, S.; Debroye, E.; Kim, C. W.; Yuan, H.; Heo, N. H.; Vanacken, J.; et al. Giant Electron–Phonon Coupling and Deep Conduction Band Resonance in Metal Halide Double Perovskite. *ACS Nano* **2018**, *12* (8), 8081–8090. https://doi.org/10.1021/acsnano.8b02936.

(35) Dahod, N. S.; France-Lanord, A.; Paritmongkol, W.; Grossman, J. C.; Tisdale, W. A. Low-Frequency Raman Spectrum of 2D Layered Perovskites: Local Atomistic Motion or Superlattice Modes? *J. Chem. Phys.* **2020**, *153* (4), 044710. https://doi.org/10.1063/5.0012763.

(36) Mauck, C. M.; France-Lanord, A.; Hernandez Oendra, A. C.; Dahod, N. S.; Grossman, J. C.; Tisdale, W. A. Inorganic Cage Motion Dominates Excited-State Dynamics in 2D-Layered Perovskites ($C_xH_{2x+1}NH_3)_2PbI_4$ ($x$ = 4–9). *J. Phys. Chem. C* **2019**, *123* (45), 27904–27916. https://doi.org/10.1021/acs.jpcc.9b07933.

(37) Dhanabalan, B.; Leng, Y.-C.; Biffi, G.; Lin, M.-L.; Tan, P.-H.; Infante, I.; Manna, L.; Arciniegas, M. P.; Krahne, R. Directional Anisotropy of the Vibrational Modes in 2D-Layered Perovskites. *ACS Nano* **2020**, *14* (4), 4689–4697. https://doi.org/10.1021/acsnano.0c00435.

(38) Gong, X.; Voznyy, O.; Jain, A.; Liu, W.; Sabatini, R.; Piontkowski, Z.; Walters, G.; Bappi, G.; Nokhrin, S.; Bushuyev, O.; et al. Electron–Phonon Interaction in Efficient Perovskite Blue Emitters. *Nat. Mater.* **2018**, *17* (6), 550–556. https://doi.org/10.1038/s41563-018-0081-x.

(39) Krylov, A. S.; Vtyurin, A. N.; Bulou, A.; Voronov, V. N. Raman Spectra and Phase Transitions in the $Rb_2KScF_6$ Elpasolite. *Ferroelectrics* **2003**, *284* (1), 47–64. https://doi.org/10.1080/00150190390204709.

(40) da Costa, A. M. A.; Geraldes, C. F. G. C.; Teixeira-Dias, J. J. C. A Raman Spectroscopic Study of Molecular Interaction in Long-Chain Primary Amines Systems. *J. Raman Spectrosc.* **1982**, *13* (1), 56–62. https://doi.org/10.1002/jrs.1250130111.

(41) Abid, H.; Trigui, A.; Mlayah, A.; Hlil, E. K.; Abid, Y. Phase Transition in Organic–Inorganic Perovskite $(C_9H_{19}NH_3)_2PbI_2Br_2$ of Long-Chain Alkylammonium. *Results Phys.* **2012**, *2*, 71–76. https://doi.org/10.1016/j.rinp.2012.04.003.

(42) Maloney, A. G. P.; Wood, P. A.; Parsons, S. Competition between Hydrogen Bonding and Dispersion Interactions in the Crystal Structures of the Primary Amines. *CrystEngComm* **2014**, *16* (19), 3867–3882. https://doi.org/10.1039/C3CE42639D.

(43) Palummo, M.; Berrios, E.; Varsano, D.; Giorgi, G. Optical Properties of Lead-Free Double Perovskites by Ab Initio Excited-State Methods. *ACS Energy Lett.* **2020**, *5* (2), 457–463. https://doi.org/10.1021/acsenergylett.9b02593.

(44) Zhang, W.; Hong, M.; Luo, J. Halide Double Perovskite Ferroelectrics. *Angew. Chem. Int. Ed.* **2020**, *59* (24), 9305–9308. https://doi.org/10.1002/anie.201916254.

(45) Soler, J. M.; Artacho, E.; Gale, J. D.; García, A.; Junquera, J.; Ordejón, P.; Sánchez-Portal, D. The SIESTA Method for *Ab Initio* Order- *N* Materials Simulation. *J. Phys. Condens. Matter* **2002**, *14* (11), 2745–2779. https://doi.org/10.1088/0953-8984/14/11/302.

(46) Giannozzi, P.; Baroni, S.; Bonini, N.; Calandra, M.; Car, R.; Cavazzoni, C.; Ceresoli, D.; Chiarotti, G. L.; Cococcioni, M.; Dabo, I.; et al. QUANTUM ESPRESSO: A Modular and Open-Source Software Project for Quantum Simulations of Materials. *J. Phys. Condens. Matter* **2009**, *21* (39), 395502. https://doi.org/10.1088/0953-8984/21/39/395502.

(47) Glazer, A. M. The Classification of Tilted Octahedra in Perovskites. *Acta Crystallogr. B* **1972**, *28* (11), 3384–3392. https://doi.org/10.1107/S0567740872007976.